%% file: main.tex
\documentclass[manuscript]{acmart}

\AtBeginDocument{%
  \providecommand\BibTeX{{%
    \normalfont B\kern-0.5em{\scshape i\kern-0.25em b}\kern-0.8em\TeX}}}

\usepackage{braket}
\usepackage{mathtools}
\usepackage{amsmath}
\usepackage{xcolor}
\usepackage{url}
\usepackage{textgreek}
\usepackage{graphicx}
\usepackage{subcaption}
\usepackage{siunitx}

\renewcommand\footnotetextcopyrightpermission[1]{} % removes footnote with conference information in first column
\begin{document}

%%
%% The "title" command has an optional parameter,
%% allowing the author to define a "short title" to be used in page headers.
\title{Quantum Computing: An Overview Across the System Stack}

%%
%% The "author" command and its associated commands are used to define
%% the authors and their affiliations.
%% Of note is the shared affiliation of the first two authors, and the
%% "authornote" and "authornotemark" commands
%% used to denote shared contribution to the research.
\author{Salonik Resch}
\email{resc0059@umn.edu}
\orcid{1234-5678-9012}
\affiliation{%
  \institution{University of Minnesota}
  \streetaddress{200 Union St SE  }
  \city{Minneapolis}
  \state{Minnesota}
  \postcode{55455}
}

\author{Ulya R. Karpuzcu}
\affiliation{%
  \institution{University of Minnesota}
  \streetaddress{200 Union St SE  }
  \city{Minneapolis}
  \state{Minnesota}
  \postcode{55455}
}

%%
%% By default, the full list of authors will be used in the page
%% headers. Often, this list is too long, and will overlap
%% other information printed in the page headers. This command allows
%% the author to define a more concise list
%% of authors' names for this purpose.
\renewcommand{\shortauthors}{Resch, et al.}

%
% The abstract is a short summary of the work to be presented in the article.
\begin{abstract}
Quantum computers, if fully realized, promise to be a revolutionary technology. As a result, quantum computing has become one of the hottest areas of research in the last few years. Much effort is being applied at all levels of the system stack, from the creation of quantum algorithms to the development of hardware devices. The quantum age appears to be arriving sooner rather than later as commercially useful small-to-medium sized machines have already been built. However, full-scale quantum computers, and the full-scale algorithms they would perform, remain out of reach for now. It is currently uncertain how the first such computer will be built. Many different technologies are competing to be the first scalable quantum computer.
\end{abstract}

%%
%% The code below is generated by the tool at http://dl.acm.org/ccs.cfm.
%% Please copy and paste the code instead of the example below.
%%

%
% Keywords. The author(s) should pick words that accurately describe the work being
% presented. Separate the keywords with commas.

%%
%% This command processes the author and affiliation and title
%% information and builds the first part of the formatted document.
\maketitle

\input{content/introduction}
\input{content/basics.tex}
\input{content/whatspossible}

\input{content/whatsbeendone}
\input{content/future}

\section*{Acknowledgements}
We thank Yipeng Huang and Brett Heischmidt for extensive feedback and helpful discussion. 

%%
%% The next two lines define the bibliography style to be used, and
%% the bibliography file.
\nocite{*}
\bibliographystyle{unsrt}
\bibliography{ref}

\end{document}

%% file: content/introduction.tex
\section{Introduction}
So what's all the hype about? Much of the excitement comes from simply how bizarre the quantum realm is. Quantum phenomena are counter intuitive and the invention of quantum mechanics changed the way we see the world. Despite being very strange, quantum mechanics is the most experimentally accurate and consistent scientific theory to date \cite{griffiths2003consistent,quantphil,nielsen2000quantum}. Quantum computation is the act of using quantum mechanics to perform computation. Quantum phenomena, such as superposition and entanglement, enable quantum computers to perform operations that have no counterpart in the classical world. These are powerful mechanisms that give quantum computers a key advantage over classical computers. Significantly, there are important scientific problems  for which quantum computers are the only known means by which to solve. Classical algorithm solutions exist, but either by excessive time (millions of years) or memory usage (more bytes of storage than atoms in the universe), are rendered entirely impractical. Thus, the creation of a full-scale quantum computer would be a revolutionary achievement. 

Quantum computers can be implemented with a variety of physical technologies, such as trapped ions, superconductors, or photons. There are advantages to each approach. However, in each case, they are very hard to build. A common issue for each approach is quantum noise. Quantum mechanical states are extremely fragile and require near absolute isolation from the environment. Such conditions are hard to create and typically require temperatures near absolute zero and shielding from radiation. Thus, building quantum computers is expensive and difficult. The challenge increases dramatically with increasing size (number of qubits and the length of time they must be coherent) and thus only small to medium scale computers have been built so far. 

Achieving full-scale computers is going to require a lot of work from many different fields, such as the design of quantum algorithms and error correcting codes, the architecture design of the computer itself, and the development of more reliable quantum devices. The difficulty of these challenges have led critics to claim quantum computing is a long way off, potentially hundreds of years away from full-scale implementation. Some have even suggested it is not possible, even in theory \cite{kalai2014gaussian, caseagainst}. However, with improving hardware, error correcting codes, and fault tolerance in target applications, the research community remains optimistic \cite{casefor,maslov2018outlook}. Much of this progress is due to intense research efforts from large technology companies, such as Google, Microsoft, and IBM. Quantum research is also becoming ever more prevalent in academia. Recently, a large amount of government funding has been devoted to quantum research, due to its potential benefits and impact on national security \cite{CNN, NQIA, EUquantum}. Additionally, numerous quantum computing start ups have appeared in the last few years. So while quantum computing is certainly still a number of years away, it is being pursued with vigor. Sooner rather than later, we will see if quantum computation becomes a revolutionary technology or falls into obscurity. This review is divided into two major sections. In Section \ref{sec:potential}, we review the potential of quantum computation and its possible applications. We summarize efforts at different levels of the system stack to provide some context on the state of the art. In section \ref{sec:scalability}, we evaluate different physical approaches to quantum computing. We collect and compare benchmarking data from various experiments and discuss critical issues to achieve scalability.

%% file: content/basics.tex
\section{Potential and Progress of Quantum Computing}
\label{sec:potential}
\subsection{Basics of Quantum Mechanics}
Classical physics accurately describes the macroscopic world. However, this does not hold at the atomic scale, where classical reasoning fails to explain strange behavior, such as in the famous double-slit experiment \cite{young1802ii,doubleslitfeynman2006qed}. Such inconsistencies led to the invention of quantum mechanics. Many implications of quantum mechanics are counter-intuitive, however they are consistently demonstrated to be true and quantum mechanics remains the best known way to describe the world \cite{griffiths2003consistent,quantphil}. In fact, classical physics appears to be an approximation of quantum mechanics at large scales \cite{jaeger2014quantum}. This is analogous to classical physics being an approximation of special relativity at sub-light speeds.

In quantum mechanics, physical properties can take on discrete values. This means that quantum mechanical systems can exist in different, distinct states. The energy levels of ions, the spin of an electron, and the polarization of a photon are all examples of such states. Quantum mechanics can be used to perform computation by assigning logical values to different states. Transitions between these states then represent logical operations. For example, with an electron spin, spin-up can be assigned logic 1 and spin-down can be logic 0. These discrete states, and corresponding discrete values, allow for digital computation \cite{nielsen2000quantum}. If the quantum system has two states, such as electron spin, it is called a qubit. Quantum systems with more than two states are called qudits. Qubits are vastly more popular and are generally assumed, as we will for the remainder of this paper. These quantum mechanical representations of information have some powerful advantages over their classical counterparts. While quantum systems have discrete states, they can exist in multiple states simultaneously. This means a qubit can be both 0 and 1 at the same time. This property is referred to as \textit{superposition}. If a qubit is in a superposition of states, a logical operation applied to it will operate on both states simultaneously. \textit{Entanglement} is the other major advantage. When multiple qubits are entangled, it means the states of individual qubits are dependant on others. Thus, information is stored not only in each qubit, but in the relationship between them. As a direct result, the amount of information stored in a combined quantum system is exponential in the number of qubits it contains. Entanglement also enables the process of \textit{quantum teleportation}. Through this process, quantum states can be transferred between two qubits by using only classical signals, if the two qubits were previously entangled. Teleportation is a misleading name as this does not imply communication faster than light, however it does provide a good channel for long distance communication. Superposition and entanglement combine to create the main power of quantum computing, \textit{quantum parallelism}. With quantum parallelism, quantum computers can perform computation on all possible inputs at the same time. This enables quantum computers to implement algorithms that no classical computer will ever be able to. This feat is referred to as \textit{Quantum Supremacy}. While certainly true of theoretical, larger-scale quantum computers, proving that it is physically realizable has been a challenge.

%\subsection{Qubits and Algebra}
\subsection{Qubits: Fundamental Building Blocks of Quantum Computers}
Qubits are the quantum version of classical bits. Just like bits, qubits have two states and can represent information in binary format. However, qubits can exist in both states at the same time due to superposition. The information contained in a single qubit can be described by two complex numbers and is typically written as $$\alpha \ket 0 + \beta \ket 1$$ 
The coefficients, $\alpha$ and $\beta$, are the complex numbers and are commonly called \textit{amplitudes}. $\ket 0$ and $\ket 1$ are the possible states of the qubit, which are commonly referred to as \textit{kets}. The coefficients represent ``how much'' of the qubit is in state $\ket 0$ and how much it is in state $\ket 1$. They determine the probability of finding the qubit in each state when it is measured.
$$prob(0) = |\alpha|^2$$
$$prob(1) = |\beta|^2$$ 
$$|\alpha|^2 + |\beta|^2 = 1$$
The third condition is because the qubit must be in one of its two states. While the magnitudes of the coefficients determine the probabilities, it is not the whole story. Significantly, $\alpha$ and $\beta$ can be negative and contain imaginary parts. Such attributes allow there to be a phase difference between the states $\ket 0$ and $\ket 1$. While this phase difference doesn't affect the probability of measurement, it does affect the qubit state during computation and is a crucial component of most quantum algorithms.

This notation can be expanded to multiple qubits. Say there is one qubit, represented by $\alpha \ket 0 + \beta \ket 1$, and another one represented by $\gamma \ket 0 + \delta \ket 1$. These two independent qubits can be combined into a 2-qubit system. This is algebraically represented by a tensor product, $\otimes$. 
$$\left (\alpha \ket 0 + \beta \ket 1 \right ) \otimes (\gamma \ket 0 + \delta \ket 1)$$
As the two qubits are independent, the state of the system can be fully described with just two complex numbers per qubit. To represent entangled states, where amplitudes of each qubit are not independent, the kets must represent the combined system. The above state can be rewritten as
$$ \alpha \gamma \ket{00} + \alpha \delta \ket{01} + \beta \gamma \ket{10} + \beta \delta \ket{11}$$
While the qubits are still independent, note that the kets now represent the states of both qubits. The states act like a 2-bit integer, and the 4 coefficients now represent the 4 different possible states of the combined quantum system. Every time a single qubit is added, the number of possible states doubles, representing the exponential information stored in a quantum state. As the above state represents independent qubits, the coefficients can be factored back into the original form. This is not true for entangled states, where the quantum state cannot be described by specifying the state of each qubit separately. An example of an entangled state is 
$$\alpha \ket{00} + \beta \ket{11}$$
There are two qubits, but the coefficients for $\ket{01}$ and $\ket{10}$ are 0. Thus, the system must be either in state $\ket{00}$ or $\ket{11}$. This is a fascinating and counter-intuitive condition. Both qubits are in a superposition of $\ket 0$ and $\ket 1$, and yet are defined to be in the same state. If one qubit is measured, the states of both qubits will be known. This state is one of the famous Bell states, and is used in quantum teleportation.

If $n$ qubits in a quantum system, there are $2^n$ amplitudes. Amplitudes act like the variables of a quantum algorithm. Through the progress of a quantum program the magnitudes and relative phases of the amplitudes will change. This provides a lot of computational power, as while there are $2^n$ ``variables'', only $n$ qubits needs to be manipulated. The caveat is that amplitudes cannot be directly measured. When the qubits are measured, the amplitudes are forced to be either 0 or 1 and only classical information is obtained. Additionally, the quantum information is destroyed in the measurement process. Thus, the final output of a quantum algorithm is at most $n$ classical bits.  Therefore, the key to quantum algorithm design is to use quantum information during computation to create a high probability of measuring a useful classical result. This has proved to be a conceptual challenge, and the design of quantum algorithms is not straight forward. An excellent introduction to quantum mechanics and algorithms can be found in \cite{loceff2015course}.

\subsection{Relation to Classical Computing}
{The state of quantum computation has been compared to classical computing in the 1950's \cite{quantumCS}, where the fundamental building blocks existed but the full-fledged system stack did not. The authors of \cite{blume2019metrics} say that quantum technology is much closer to classical hardware of 1938. At this time it was unknown what technology was going to prevail, how such devices would be built, or how they would be integrated to construct useful machines. In either case, the development of classical computers decades ago acts as a road map for the development of quantum computers today.} Quantum devices are inherently different and bring their own challenges, resulting in tighter resource constraints \cite{quantumCS}, but many of the concepts are the same. The construction of large scale quantum computers will require the development of quantum versions of devices, architectures, languages, compilers, and layers of abstraction. These quantum versions appear to pose greater challenges, but knowledge gained from the creation of their classical counterparts will prove useful. Much of the work will be taking classical concepts and adapting them to work for quantum systems.

\subsubsection{Memory}
A fundamental and major difference between classical and quantum computing is the concept of memory. Classically, there is a natural distinction between memory and computation. Data can easily be made permanent and moved around as needed. Disk memory can store data for arbitrarily long times, which can be copied and used as input for computation. The results can also be sent back to permanent storage. The situation is quite different for quantum computers, where the quantum data typically only last as long as the program. There are a number of reasons for this. A fundamental limitation comes from the fact that a quantum state cannot be copied, according to the no-cloning theorem \cite{wootters1982single}. Thus, there is no way to create multiple copies of the same data, other than to perform the lengthy state preparation multiple times. As a result, quantum computation always operates directly on the input data. In this way, quantum computing is similar to the classical concept of processing-in-memory, where values stored in memory are used as the inputs to logic operations. However, by definition, quantum operations ``overwrite'' the data in the process. Physical limitations of quantum devices also put severe restrictions on quantum memory. Quantum states are fragile and naturally decay. Thus, it is extremely difficult to maintain quantum information for long periods of time. Even maintaining data for the length of the program, the equivalent of RAM, poses a considerable challenge. No known quantum technology can provide the equivalent of non-volatile disk storage. There have been proposals for quantum computers with separate compute and memory regions \cite{oskin2002practical,chi2007tailoring,beals2013efficient}, however data loss in the memory remains an issue \cite{ahsan2015optimization}. Some technologies have demonstrated relatively long coherence times \cite{sukachev2017silicon}, which could in theory act as a quantum memory. However, these are at most a few seconds and have yet to be integrated with computation. Additionally, modern quantum computers have not achieved high qubit counts. Hence, from a practical physical perspective, there currently is not much demand for quantum memory.

As quantum technology develops, quantum memory may become more practical. This would be desirable, as a number of proposed large-scale quantum algorithms would heavily depend on them \cite{rebentrost2014quantum}. There have been proposals for quantum RAM (QRAM), such as the \emph{Bucket Brigade} approach \cite{QRAMBucketBrigadegiovannetti2008quantum}. However, as noted by Di Matteo et. al. \cite{qRAMResourcedi2019fault}, implementing fault-tolerant QRAMs of this nature take impractical resources, well beyond what is possible in the foreseeable future. For example, an 8GB QRAM with millisecond access times would require quadrillions of qubits, or, if using only a few million qubits, the access time would be a few years. Using an alternative approach, \emph{Select Swap} \cite{SelectSwaplow2018trading}, could significantly improve performance. For example, an access time of 10 seconds could be achieved when using approximately 100,000 qubits \cite{MatteoPresentation}. While more hopeful, this is still well beyond capabilities of near term computers. Clearly, physical implementations of quantum computers need to significantly improve before QRAM can be considered as a viable option.

\subsubsection{Approximate Computing}
 A modern classical concept that has relevance for quantum computing is approximate computing. Quantum computing is probabilistic by nature, it is very common for a quantum algorithm to only have a reasonable probability of providing the correct answer \cite{schuldtext,loceff2015course}. The answer can be checked for correctness with classical computations or the quantum algorithm can be run enough times to generate a sufficient certainty in the results. However, this probabilistic nature of quantum computing is not entirely equivalent to classical approximate computing. In many quantum algorithms, if the result is wrong, it is not usable. Quantum states that are very far from the correct result will usually have some probability of being measured. In this case, the measured result is not an approximation and there is no way to recover a correct answer; the algorithm must simply be restarted. However, a recent trend has been the combination of classical optimization techniques and quantum subroutines. Prominent algorithms such as the Quantum Approximate Optimization Algorithm (QAOA) \cite{farhi2014quantum} and the Variational Quantum Eigensolver \cite{peruzzo2014variational} make use of this approach. For these algorithms, the measured results are approximations. The algorithms switch between classical optimization and quantum subroutines, and the results can be improved over multiple iterations.

\subsubsection{The Position of Quantum Technology}
It is often noted that quantum computing, in the foreseeable future, is not a direct competitor to classical computers. The general notion is that quantum computers are capable of things that classical ones are not, but typically perform relatively poor on applications that classical computers can do well. The main reason for this is that quantum operations are slow, relative to classical transistor-based operations, and are error prone. Thus, there is often not much to gain by replacing a classical computer with a quantum one. This mindset restricts quantum computing to highly specific applications and cloud based services \cite{xin2018nmrcloudq}. While this is likely to remain true, it should be noted that this is not \textit{necessarily} a fundamental limitation. Quantum computers are slow and noisy, in large part, because of how they are currently made. Depending on the underlying technology, it is possible that someday quantum computers will be nearly or just as fast as modern classical ones, and will have sufficiently sophisticated mechanisms for handling noise. If this is the case, they could replace classical computers for general purpose computing. However, such developments are \textit{many} years away, if they are possible at all. Regardless of whether quantum computation will become ubiquitous or not, quantum technology will undoubtedly be of great use. Quantum computers would be a great addition to modern supercomputing clusters, where, for example, they would greatly improve capabilities of simulations of molecules and materials \cite{FutureComputingsvore2016quantum}. Other relevant and useful applications have already been found. It is possible that the best uses of quantum hardware are currently unknown to us. For example, in addition to quantum computing, quantum metrology and quantum communication show much promise. Quantum metrology uses quantum mechanics to perform more accurate measurements that those that are possible classically \cite{metrologygiovannetti2006quantum}. Perfectly secure channels are enabled by quantum communication \cite{communicationgisin2007quantum}. Both of these fields not only are scientifically relevant, but could create highly valuable commercial products. 

\subsection{Skepticism}
Despite the enthusiasm surrounding quantum computing, it is not without its critics. And indeed, there is good reason to be skeptical. Building quantum computers has proven deceptively difficult. Despite intensive effort, quantum noise currently remains an formidable barrier. It's been estimated that more than 99\% of the computation performed by a quantum computer will be for error correction \cite{kreger2008microcoded}. Some skeptics suggest that quantum supremacy may be truly impossible due to noise inherent in quantum computing \cite{kalai2014gaussian}. It was argued in \cite{caseagainst} that the analog nature of quantum information results in the need to require too many variables, rendering it an unrealistic task. However, the authors of \cite{casefor} say that quantum computer design is a digital discipline, and suggest that scalable quantum computing is possible through modular design and error correcting codes. Many believe that it is simply a problem of engineering \cite{oskin2002practical,progressprospectsnational2019quantum}. Some remain skeptical about quantum computing, but believe other quantum technologies will be rewarding \cite{aspect2004john}. The next few years will tell.

%% file: content/whatspossible.tex
\subsection{Potential and Motivation}
%\begin{itemize}
%    \item Full-scale quantum algorithms
%    \item All possible applications
%    \item What's needed for these to work
%\end{itemize}
Quantum computers are exciting and appealing because they operate in a completely different way than modern computers. But in more a practical sense, quantum computing is worth pursuing because of the advantages it offers in relevant scientific and commercial applications. The applicability of quantum computing almost always follows the same pattern. 

\begin{enumerate}
    \item There is a known problem which has great rewards if solved.
    \item Classical algorithms exist to solve such problems, but they take excessive time or memory to complete, rendering them impractical.
    \item It is believed that no reasonable classical solution exists.
    \item There is a quantum algorithm which can solve the problem under practical time and resource constraints.
    \item Unfortunately, quantum computers are not yet mature enough to effectively run the algorithm.
\end{enumerate}

The main idea is that, historically, it is the algorithms that have motivated the development of full scale quantum computing. The potential of a large-scale quantum computer is already known, despite the fact that it doesn't exist yet. This makes the research somewhat unique as the end goal is clear, but the technology must be developed in order to get there. Thus, large scale algorithms provide the best indicators of the full potential future of quantum computing\footnote{Different methods of utilizing quantum information may be developed which will change how quantum computers are used. Thus, quantum algorithms represent the distant future of quantum computing as we see it today, which may change as the technology develops. This trend has already been seen as useful applications have been developed for small-to-medium sized quantum computers.}.

Peter Shor's factorization algorithm \cite{shor1999polynomial} is likely most responsible for generating interest in quantum computing. The algorithm efficiently factors a very large number into its constituent primes. Its based on the relationship between the given number to factor and the periodicity of a function defined by it \cite{loceff2015course}. Quantum parallelism enables a quantum oracle (set of quantum gates) to evaluate every input value of the function simultaneously. The quantum fourier transform (QFT), also developed by Shor, enables a measurement in the \textit{frequency} domain, which is likely to report a number close to the period. Classical post processing, including the continued fractions and Euclidean algorithms, can be used to extract the exact period and find the constituent primes. A quantum computer running Shor's algorithm would have a completion time on the order of seconds to minutes. This is significant because RSA, the encryption scheme responsible for internet security, relies on the fact that such factoring is a hard problem. Classical best-known approaches take years (possibly billions of years) to complete \cite{blueprint}\cite{postquantumcrypt}. If one could factor large numbers quickly, one could also break internet security.  Thus, Shor's algorithm reveals that the existence of a quantum computer would have an immediate impact and has become the \textit{prime} example of the potential of quantum computing. The QFT is very similar to the classical discrete time fourier transform (DFT), which is currently one of the mostly valuable and widely used algorithms. Unfortunately, the QFT cannot be used in the same manner that the DFT is \cite{loceff2015course}. 

Quantum chemistry is another major application that will benefit from the development of quantum computing \cite{QuantumChemistryAolson2017quantum,QuantumChemistryBmcardle2018quantum,QuantumChemistryCcao2018quantum}. It is possibly the most useful known application. Classical simulation of quantum chemistry is of interest in many fields, including chemistry, condensed matter physics, materials science, biophysics, and bio chemistry; but is limited by the exponential increase in required resources with problem size \cite{lanyon2010towards,whitfield2011simulation}. Thus, simulations are limited to finding the ground state energy of extremely small collections of molecules \cite{peruzzo2014variational,thogersen2004coupled}. Quantum computing offers a natural solution.  A quantum computer's resources increase exponentially with the number of qubits it possesses, and could use Quantum Phase estimation \cite{lloyd1996universal} to efficiently solve the problem \cite{wang2018generalised}. For a quantum computer, the problem of quantum chemistry would scale polynomially \cite{lanyon2010towards,whitfield2011simulation,nielsen2000quantum,dumitrescu2018cloud}. This could speed up simulations that are presently being run and enable simulation of larger systems and excited energy levels. 

Grover's search algorithm \cite{grover1996fast} provides a quadratic speedup in uninformed search. While this is not as a significant of a speedup, it provides a fundamental and powerful example of quantum computing's potential. Performing search in sub-linear time is something clearly impossible in classical computing.

%% file: content/whatsbeendone.tex
\subsection{Progress Towards the Quantum Era}
 While the tasks ahead are daunting, the enticing rewards of quantum research has generated an extreme amount of recent progress. Quantum has quickly become one of the hottest topics and there are many developments at all levels of the system stack. A trend that has emerged is the use of noisy intermediate-scale quantum (NISQ) devices \cite{preskill2018quantum}. While advances towards full-scale quantum computers are on their way, it's been found that current quantum computers have potential of being useful, despite being small and noisy. Thus, finding applications for these attainable quantum devices has become a significant focus, in addition to the drive towards full scale computers.

\subsubsection{Fully-Functional Computers}
Google \cite{google}, IBM (Q Experience 5,16, Q17 \cite{IBM1617} Q20 \cite{IBM2050}), Intel \cite{intel17}, and Rigetti \cite{rigetti}
have all built quantum chips \cite{castelvecchi2017quantum}, all of which are superconducting architectures. Vancouver based company D-Wave \cite{rose2007introduction} has built adiabatic quantum computers. IBM had the highest qubit count with 50 at the end of 2017 \cite{IBMmit}. Google announced its 72-qubit quantum computer in March 2018, Bristlecone, which is the highest number of qubits to date \cite{bristlecone}. IBM announced System One \cite{IBMqsys1} in early 2019. It has been designed specifically for scientific and commercial use, intended applications include modelling financial data and optimizing fleet operations for deliveries. In late summmer 2019, Google demonstrated a 52-qubit superconducting chip, named Sycamore, which they claim has demonstrated quantum supremacy \cite{arute2019quantum}. IonQ \cite{ionq}, a  quantum computing startup company in Maryland has built the first ion-trap based quantum computer for commercial use \cite{ionqphysicsworld}. 
Given that nearly all competitive quantum computers are superconductor based, the success of IonQ's machine came as a bit of a surprise \cite{ionqphysicsworld}.  A large number of startups have been founded with intentions of building quantum computers, a comprehensive list of which can be found at \cite{startups}.

\subsubsection{Software}
Quantum computers run quantum programs. Thus, there is need for quantum languages and compilers in order to create and build quantum algorithms. The circuit model of quantum computing, which is the most amenable from an algorithm perspective \cite{dunjko2018machine}, consists of sequence of quantum gates (unitary operations). Thus, quantum languages and compilers should facilitate the conversion from high level descriptions to individual gates and the control signals necessary to perform them. A number of languages and compilers exist. qcl \cite{qcl} and Scaffold \cite{abhari2012scaffold} are quantum languages based on C and \cite{bettelli2003toward} is on C++. $Liqui\mid> $ \cite{wecker2014liqui} is a software architecture and toolsuite from Microsoft, which is written in F\#. Additionally, Q\# \cite{svore2018q} is a domain specific language which is part of Microsofts Quantum Development Kit \cite{microsoft}.

Along with software comes the issues of the instruction set architecture (ISA), which are the operations that the quantum computer can understand, and the corresponding compilers that generate them. Quantum compilation is complex as there are many complicating factors and each can significantly affect the performance of the final program \cite{Softwaremethodologyhaner2018software}. A number of ISAs have been suggested such as a von Neumann architecture-based virtual-instruction set \cite{balensiefer2005evaluation}, quantum physical operations language (QPOL) \cite{svore2006layered}, Quil \cite{smith2016practical}, and OPENQASM \cite{cross2017open}. Quipper \cite{green2013quipper} is a quantum compiler implemented in Haskell and Scaffcc \cite{javadiabhari2014scaffcc} is one implemented on LLVM. It was noted in \cite{fu2017experimental} that many of these ISAs, and corresponding compilers, focus on intermediate representations and do not consider low-level constraints involved in interfacing with a quantum computer. Their approach was to break down quantum instructions into micro-instructions, which are co-designed with an architecture. The corresponding ISA is called eQASM \cite{fu2018eqasm}.  The authors of \cite{Softwaremethodologyhaner2018software} present a methodology for compilation and discuss required abstractions. They noted it can be beneficial to have multiple levels of compilers which generate multiple quantum intermediate representations (QIR). This enables code reuse at higher levels, which are agnostic of the gate set and hardware. Additionally, optimization can be done separately at different layers, which allows high level optimizations to be combined with hardware or machine specific optimizations. The authors of \cite{shi2019optimized} proposed a compilation technique where individual quantum instructions are aggregated into larger operations. These aggregated instructions enable better optimization of the required control pulses, which reduces the latency of the program. A number of recent works have adapted compilers to mitigate the effects of physical errors on near-term machines. Two common objectives are scheduling operations on the most reliable hardware and reducing overhead for 2-qubit gates. Most computers require qubits to be adjacent when performing 2-qubit gates. Moving qubits into the appropriate positions is costly as this often requires high-latency and error prone swap operations. The authors of \cite{zulehner2018efficient} proposed an efficient qubit mapping method for IBM QX architectures which demonstrated a significant improvement over IBM's own mapping solution. A compiler proposed in \cite{murali2019noise} maps quantum algorithms to run on the IBMQ16 \cite{IBM1617}. The underlying hardware is frequently calibrated and the qubit coherence times and gate error rates change daily. The compiler is aware of these changes and optimizes to qubit placement to maximize the probability of successful program completion. The error rates on the IBM-Q20 \cite{IBM2050} were studied over a long period of time in \cite{tannu2018case}, where they developed variation aware qubit movement and allocation policies. A more general version of this problem is mapping logical qubits onto physical ones for arbitrary topologies. Two recent works, SABRE \cite{li2019tackling} and t$\ket{ket}$ \cite{cowtan2019qubit}, address this problem. The goal of each is to reduce overhead for 2-qubit operations by finding optimal initial placement and qubit movements. A recently propose toolflow, TriQ \cite{TriQmurali2019full}, uses real-system measurements (from IBM, Rigetti, and IonQ) to optimize compilation for these systems. TriQ was able to significantly improve the success rate of the implemented quantum algorithms.

Numerous quantum computing related add-on software packages exist, such as QuTip \cite{qutip,johansson2012qutip} and Pennylane \cite{bergholm2018pennylane} for python, QCSimulator \cite{QCSimulator} and QuantumOps \cite{QuantumOps} for R \cite{R}, QUBIT4MATLAB for MATLAB \cite{MATLAB:2010}. An overview of gate-level quantum software platforms can be found in \cite{larose2018overview}. ProjectQ \cite{steiger2018projectq} is an open source software effort based in python. It is an attempt to consolidate efforts of quantum researchers by creating a shared platform which facilitates code sharing and re-use. The result would be similar to the machine learning community which produces a substantial python library. It enables the simulation of quantum algorithms and can connect to the IBM quantum experience \cite{ibmexperience} cloud service and run quantum algorithms on real hardware.

It is not reasonable to expect a programmer to create a program that is entirely correct. With high certainty, flaws will be present due to human error. Accounting for this limitation is necessary, and hence, effective debugging is an essential component of writing software. Debugging is more challenging for quantum programs as they not only inherit the difficulties of designing classical algorithms but introduce additional complexity and are harder to analyze \cite{huang2018qdb}. There are two general methods to overcome these challenges. One is creating formal definitions to prevent the introduction of mistakes. QWIRE \cite{QWIREpaykin2017qwire} separates definitions of quantum operations from a higher level host language, which prevents the host language from breaking the laws of quantum mechanics. The authors of \cite{RobustnessAnalysishung2019quantitative} develop methods for reasoning about quantum errors and found a metric to characterize the ``distance'' between correct programs and ones with errors. Another approach, perhaps more familiar to programmers and engineers, is conventional assertion and debugging techniques. Huang et. al. \cite{huang2019statistical} proposed strategies to debug quantum programs. Such techniques are commonly used for classical programs, however the situation is more difficult for quantum programs. Quantum systems can have huge state spaces, making them difficult to describe and validate. Additionally, measuring states in the middle of a program will destroy the intermediate results, requiring a full restart \cite{huang2018qdb}. In \cite{huang2019statistical}, the authors propose using ``quantum breakpoints'' which are cleverly inserted to check for key properties during the run of a program. These include checking if qubits are in the correct superposition of states and if qubits are properly entangled. These checks can detect mistakes in the code introduced by the programmer or, when running the program on a real machine, can help identify where physical errors have corrupted the state.

\subsubsection{Algorithms}
There are a few famous quantum algorithms which have demonstrated the power of quantum computing. Shor's algorithm \cite{shor1999polynomial} uses the QFT to factor large prime numbers. Grover's algorithm \cite{grover1996fast} can perform sub-linear search. Simon's  \cite{simon1997power} and the Deutsch-Jozsa \cite{deutsch1992rapid} algorithms not useful for practical purposes but demonstrate quantum potential. Simon's finds the period of a function and the Deutsch-Jozsa determines if a function is constant or balanced in one iteration. There are a growing number of other useful algorithms \cite{montanaro2016quantum}, such as quantum versions of financial portfolio optimization \cite{rebentrost2018quantum}, random walk \cite{childs2002example,kempe2005discrete,aharonov2001quantum}, and Blockchain \cite{tessler2017bitcoin,jogenfors2016quantum,ikeda2018qbitcoin,kiktenko2018quantum,rajan2018quantum}. A comprehensive list can be found in \cite{quantumalgorithms}. Approximation algorithms have been particularly intriguing. A main benefit is that they can be run on near-term, error prone quantum computers \cite{wang2018generalised,aspuru2005simulated}. They consist largely of classical computations with quantum subroutines. The process involves steering an entangled state towards a target state that minimizes a cost function, which can be done via variation of quantum gate parameters \cite{moll2018quantum,wecker2015progress}. Two of the most prominant are the Quantum Approximation Optimization Algorithm \cite{farhi2014quantum} and the Variational Quantum Eigensolver \cite{peruzzo2014variational}.

The Quantum Approximate Optimization algorithm (QAOA) \cite{farhi2014quantum} was proposed in 2014. This was exciting as it appeared to be a generalizeable algorithm which could work on near term quantum computers \cite{shorlec}, and is a very strong candidate to demonstrate quantum supremacy. Though it is still unknown how widely applicable it is. At a high level, classical input conditions called \textit{clauses} are specified, and the algorithm increases the amplitudes of quantum states that satisfy the clauses.  A useful circuit level explanation of the algorithm is provided in \cite{wang2018introduction}. The original algorithm was proposed showing how it could solve max-cut, the problem of splitting a graph to maximize the number of connections between the two halves. QAOA was used by the authors of \cite{anschuetz2018variational}, who proposed variational quantum factoring (VQF) as an alternative to Shor's algorithm. It uses QAOA to find the ground state of an Ising Hamiltonian, which is a problem that can be mapped to factoring \cite{anschuetz2018variational,factoringasoptimization}. Previously, this was attempted with quantum annealing or simulated adiabatic evolution \cite{schaller2007role,jiang2018quantum,peng2008quantum,xu3726quantum}.

The authors of \cite{peruzzo2014variational} introduced the variational quantum eigensolver (VQE) which can be used to find eigenvalues of large matrices and is useful in simulations \cite{pgrove,grove}. The quantum phase estimation (QPE) algorithm is useful for quantum chemistry but has coherency requirements that are impractical on modern quantum computers. VQE can be used as a substitute for QPE and has become popular due to its low resource requirements. It has been used a number of times for applications in quantum chemistry \cite{yung2014transistor,kandala2017hardware,romero2018strategies,dumitrescu2018cloud,moll2018quantum}. Additional theory related to VQE can be found in \cite{mcclean2016theory}. There is a concern about the number of samples required by VQE, which can lead to excessive runtime \cite{wecker2015progress}. The authors of \cite{wang2018generalised} proposed a new algorithm, \textalpha-VQE, which attempts to compromise between the larger number of samples required by VQE and the long coherence time required for QPE. %The authors of \cite{romero2018strategies} studied using the VQE to simulate molecular energies using the unitary coupled cluster ansatz. They found they could reduce the circuit depth without introducing significant accuracy loss. \cite{dumitrescu2018cloud} simulated the deuteron binding energy using the the VQE algorithm on the IBM QX5 and Rigetti 19Q quantum chips.

There's been great interest in whether quantum computation can be effective at machine learning applications. A number of quantum machine learning algorithms have been developed \cite{wossnig2018quantum,olivares2018measurement,wiebe2012quantum,lloyd2014quantum,kimmel2017hamiltonian,rebentrost2014quantum,lloyd2016quantum,dridi2015homology,rebentrost2018quantum2,schuld2016prediction,brandao2017quantum,rebentrost2016quantum,panella2011neural}. Approaches vary from translating classical machine learning models into quantum algorithms to creating new models based on the working principles of quantum computers \cite{schuldtext}. Deep neural networks (DNNs) are of particular interest due to the incredible success of deep learning in classical neural networks. Unfortunately, neural networks seem to be an unnatural fit as they require non-linearity to solve complex problems, which contradicts with the inherent linearity of quantum computation \cite{schaller2007role}. Measurements of quantum states can introduce non-linearity \cite{chen2018universal}, however this does destroy quantum information in the process \cite{schuld2014quest}. An interesting recent approach was the quantum classifier \cite{schuld2018circuit} which, similar to optimization algorithms, uses a combination of quantum and classical computations. The network is run many times with repeated measurements of the output, which introduces non-linearity. The results of the measurement are used with gradient descent to update classical parameters, which determine the quantum operations. Hopfield networks seem to be more compatible with quantum computing \cite{schuldtext}, and have gotten some attention in quantum research \cite{rebentrost2018quantumHopfield}. However, they solve a different set of problems than DNNs, commonly pattern matching and content addressable memory. Support vector machines (SVM) are another machine learning application that may provide an ideal path for integrating non-linearity with quantum computation \cite{schuldtext}. The authors of \cite{SVM} proposed a quantum SVM which uses quantum matrix inversion \cite{MatrixInversion} to solve the least-squares formulation \cite{LeastSquaresSVM} of the SVM problem. The implementation relies on quantum oracles, possibly implemented with quantum RAM \cite{QRAM}, to create quantum superpositions of input data. Another SVM proposal uses classical input data \cite{havlivcek2019supervised}. A review of the quantum approaches to neural networks can be found in \cite{schuld2014quest} and comprehensive reviews of quantum computation's applicability to machine learning can be found in \cite{dunjko2018machine,schuld2015introduction,biamonte2017quantum,schuldtext}.

\subsubsection{Network}
Quantum entanglement allows for the teleportation of quantum states across large distances. This enables the possibility of a large scale quantum network, which could act as something of a quantum internet. Such a network is of great interest as quantum encryption is unbreakable, even in theory \cite{bennett1984quantum}, allowing for absolute security. Quantum encryption would serve the same purpose on the quantum network as RSA currently does for the modern internet. Quantum communication was a key area in the European Union flagship on quantum technology \cite{simon2017towards,EUquantum}. China has taken the initiative and launched a low-earth satellite, called \textit{Micius}, which has demonstrated direct entanglement across 1,200 Km \cite{yin2017satellite,simon2017towards}. The satellite was also used to distribute entangled photon pairs to two different locations on earth and later was used to perform quantum key distribution between the two sites \cite{liao2018satellite}. Additionally, quantum teleportation over 30km was achieved via optical fibre network in \cite{sun2016quantum}. 

{While the concept of a quantum network is most associated with the quantum internet, it can also be an approach to achieving scalability for an individual quantum computer \cite{van2016path}. It is the concept of distributed computing, where the key idea is to create modular units that are relatively robust.  A large scale computer can then be built by adding additional units, which are connected via a network. A significant amount of noise can be tolerated in the network if the individual units are of high quality \cite{nickerson2013topological}. The principles used for designing a distributed system apply whenever there are both fast ``local'' and slower ``long distance'' operations, which is applicable for both monolithic quantum computers and quantum networks \cite{van2006distributed}. Distributed systems were explored in \cite{nickerson2013topological,van2006distributed}. Nodes for such networks, based on NV-centers in diamond, were demonstrated in \cite{reiserer2016robust,kalb2017entanglement,humphreys2018deterministic}. }

%The authors of \cite{nickerson2013topological} developed a noisy-network scheme which could tolerate errors in the network up to 10\%. Individual units, cells, contained 4 ancilla qubits and 1 data qubit. Multiple cells were connected in a GHZ state. An intra-cell error rate of 0.82\% was sufficient to enable correct operation, significantly higher than previous approaches. It was noted in \cite{van2006distributed} that the principles use for designing a distributed system apply whenever there are fast "local" operations and slower "long distance" operations, which is applicable for both monolithic quantum computers and quantum networks.

%The authors of \cite{humphreys2018deterministic} were able to achieve entanglement generation rates between diamond spin qubit nodes which exceeded their decoherence rates. Providing entanglement rates higher than the decoherence rate is essential for creating future, large scale quantum networks. 

%% file: content/future.tex
\section{Quantum Technology Scalability}
\label{sec:scalability}
In this section, we discuss different quantum technologies and their potential to realize scalable quantum machines. There is a wide array of approaches and a complex and rich trade-off space. It is still unknown which technology will prove to be the most effective \cite{QuantumComputersladd2010quantum}. 

\subsection{Error Correction}
When discussing the scalability of quantum technology it is necessary to understand the role of quantum error correction. Error correction is a necessary component of all computing systems. However, for quantum computers it takes center stage. Noise is essentially inevitable in quantum computation, and it's the main reason large scale computers have not been built yet \cite{polian2015design}. Any interaction with the environment will modify or destroy a quantum state. As a result, quantum states can be corrupted during manipulation and will decay over time. A remarkable response to this has been the invention of error correcting codes. The basic idea is the same as classical error correction; some form of duplication is created, which allows the state to be restored if it is only partially corrupted. Effectively, the quantum information is ``spread out'' and is less susceptible to a single disruption. A significant difference for quantum error correction is that the states cannot be measured. Thus, traditional error correcting schemes do not work. Instead, extra qubits (called ancilla) are interacted with the qubits holding the state. The ancilla qubits are then measured, the results of which will determine a set of actions to perform on the original qubits to restore the desired state. Interestingly, quantum errors are analog in nature, but the act of error correction digitizes the error \cite{metodi2006quantum,casefor}. Therefore, a small set of operations is capable of extremely precise correction. The simplest code that provides bit flip and phase errors is the Shor code \cite{shor1995scheme}, which uses 9 physical qubits to encode one logical qubit. The Steane code \cite{steane1996error} uses 7 qubits and has a recursive encoding, allowing there to be multiple levels of error correction \cite{noisy}. Due to these error correcting codes, quantum computing can proceed in spite of the inevitable quantum noise. Unfortunately the overhead for quantum error correction can be quite high, and it is highly dependant on the physical error rates \cite{oskin2002practical}. The architecture can help with this problem, as it can be designed to be efficient at the frequently performed error correcting operations \cite{HardwareECtannu2017taming,metodi2006quantum}. However, the overhead can easily become unmanageable even with modest error rates. For example, surface codes provide sufficient protection for error rates seen on modern devices, close to 1\% \cite{wang2010quantum,fowler2012towards,SurfaceCodeinSCjavadi2017optimized}. However, thousands of physical qubits are needed for just one logical qubit \cite{fowler2012surface}.  Thus, while some quantum noise can be tolerated, it remains the most challenging obstacle to building quantum computers. A introduction of quantum error correcting codes is provided in \cite{ErrorCorrectionIntroductiongottesman2010introduction}.

While error correction remains impractical for modern computers, it has been demonstrated experimentally\cite{cory1998experimental,chiaverini2004realization,schindler2011experimental}. The authors of \cite{ask1errorcorrection} performed an experiment which demonstrated an impressive physical implementation which was capable of arbitrary error correction. Seven physical qubits, implemented with trapped ions, were used in a 2-dimensional topological color code to encode a single qubit. This enabled the correction of a single bit flip, phase flip, or combination of both on any of the individual physical qubits. Additionally, they were able to successfully apply single qubit operations on the encoded state.

\subsection{Quantum Technology and Potential for Scaling}
Quantum computation can be abstracted away from the physical processes that perform it. There are numerous physical methods to perform the same logical quantum operations, each of which can be quite distinct \cite{QuantumComputersladd2010quantum}. The essential components are quantum superposition (qubits being in multiple states at once) and entanglement (the states of multiple qubits being dependent; cannot be described individually). Quantum computation has been shown to be possible using molecular magnets \cite{leuenberger2001quantum}, NMR spectroscopy \cite{cory1997ensemble,jiang2018experimental,xin2018nmrcloudq}, photons \cite{kok2007linear,milburn2009photons}, non-Abelian anyons \cite{kitaev2003fault,nayak2008non,alicea2011non}, trapped ions \cite{cirac1995quantum,ballance2016high,schafer2018fast,gaebler2016high,harty2014high,linke2017experimental}, Quantum Dots \cite{loss1998quantum,nichol2017high,yoneda2018quantum,kawakami2016gate,veldhorst2015two}, and superconductors \cite{clarke2008superconducting,devoret2013superconducting,barends2014superconducting,rigetti2012superconducting,lucero2008high,chow2009randomized,larsen2015semiconductor,ibmexperience,EngineerGuidetoSCkrantz2019quantum}. The physics describing these systems, and  the specific engineering challenges associated with each of these technologies, would be quite different. However, the goal of each is the same and some general principles apply to all approaches. According to \cite{nielsen2000quantum,divincenzo2000physical}, there are 4 conditions which any potential quantum technology must provide.
\begin{enumerate}
    \item Robustly represent quantum information
    \item Perform a universal family of unitary transformations
    \item Prepare a fiducial initial state
    \item Measure the output result
\end{enumerate}
These items have come to be known as the ``DiVincenzo criteria''. The first item is simply the ability to reliably store and use information, a concept which is shared by classical computers. However, this task is much harder for quantum computers due to decoherence. The second item is the same concept as a universal set of operations but with the additional constraint that the operations be unitary {(which only allows linear transformations)}. NAND comprises a universal set for classical computers. The Hadamard, CNOT, and $\pi/8$ gates are one example of a universal set of quantum operations \cite{nielsen2000quantum}, however the chosen set will depend on the underlying technology \cite{linke2017experimental}. The last two items refer to the classical creation and measurement of states. As humans are classical beings, a quantum computer is not of much use if the quantum operations it performs cannot be applied to classical data or produce classical results. Thus, the quantum system must be able to interact with the classical world in a meaningful way. In summary, a quantum computer must robustly perform arbitrary quantum operations and be capable of being classically controlled and measured. 

The list does not mention scalability, however it is nearly as critical as the other criteria. While interesting from a scientific perspective, a small quantum computer is not of much practical use. Even if it exceeds in all other metrics, a 5-qubit quantum computer isn't going to take on any real world applications. So the question is how to build a machine which is capable of all four criteria and is expandable to practical sizes. There are many factors at play and each physical approach to quantum computations has its own strengths and weaknesses. The characteristics of the underlying technology determine the nature and the level of difficulties in designing the architecture. While intimately related, scalability metrics for physics and engineering perspectives can be roughly divided. {From a physics (device) perspective, important metrics for scalability include coherence time, gate latency, gate fidelity, and mobility. For an engineering perspective, topology (qubit connectivity), maturity, ease of fabrication, control, and integration are important.} Each of these metrics are dependant on others and one cannot consider them individually \cite{linke2017experimental}. While each metric can't provide a full picture, it is useful to understand the general effect they have on the architectural design of a quantum computer.

\subsubsection{Physics}
\begin{itemize}
    \item \textbf{Coherence Time} is a measure of typically how long quantum states that represent qubits remain coherent. Longer times are preferable. This gives one more time to complete a quantum operation, allowing more operations (a deeper quantum circuit) to take place in a given algorithm. By the same token, error correction will have to be applied less frequently, creating a lower overhead. 
    \item \textbf{Gate Latency} is how long it takes to perform a single quantum operation. A lower latency has a similar effect to longer coherence time. The shorter gate operations are, the more that can be performed before decoherence of the quantum state occurs. Latency is highly dependent on the physical technology but is also determined by the specific methods used to implement it. There is typically an optimal latency which introduces the least amount of noise.
    \item \textbf{Gate Fidelity} is how likely a gate will be performed without introducing error. It is why coherence time and gate latency are not linearly related. While lower gate latency allows more operations to occur within the typical coherence time, these gates introduce more opportunities for decoherence. As a metric, fidelity is a simplification of the complex quantum error models \cite{e1,e2,e3,e4,e5,e6,e7,e8,e9}. Fidelity for quantum operations are substantially lower than in classical computing, often less than 99\%. 
    \item \textbf{Mobility} is whether qubits can physically be moved or not. Superconducting circuits cannot be moved, however ions in ion traps can be transported through a vacuum, and photons never stop moving. While mobility isn't a metric that can be improved as it is a characteristic of the physical device, it can impact how information is transferred between qubits and how entanglement is created. Thus, it is a factor to consider in evaluating potential implementations for computing. 
\end{itemize}

\begin{table}
\resizebox{1.0\linewidth}{!}{
\centering
\begin{tabular}{|c||c|c|c|c|c|c|}
    \hline
     %Technology & Coherence Time & Gate Latency (s) (\b{single}/\r{multi}) & Gate Fidelity \%  (\b{single}/\r{multi}) & Mobile\\
    Technology & Coherence Time (s) & 1-Qubit Gate Latency (s) & 2-Qubit Gate Latency (s) & 1-Qubit Gate Fidelity (\%) & 2-Qubit Gate Fidelity (\%) & Mobile\\
     \hline
     \hline
     Ion Trap  & 
     0.2 \cite{ballance2016high} - 0.5 \cite{linke2017experimental} & 
     1.6e-6 \cite{schafer2018fast} - 2e-5 \cite{linke2017experimental} &
     5.4e-7 \cite{schafer2018fast} - 2.5e-4 \cite{linke2017experimental} &
     99.1 \cite{linke2017experimental} - 99.9999 \cite{harty2014high} &
     97 \cite{linke2017experimental} - 99.9 \cite{ballance2016high} &
     YES \\ 
     \hline
     Superconductors  &
     7.0e-6 \cite{chen2014qubit} - 9.5e-5 \cite{rigetti2012superconducting} &
     2.0e-8 \cite{fu2017experimental,chow2009randomized,barends2014superconducting} - 1.30e-7  \cite{ibmexperience,linke2017experimental} &
     3.0e-8 \cite{chen2014qubit} - 2.5e-7 \cite{ibmexperience,linke2017experimental} &
     98 \cite{lucero2008high} - 99.92 \cite{barends2014superconducting}  &
     96.5 \cite{ibmexperience,linke2017experimental} - 99.4 \cite{barends2014superconducting} &
     NO \\
     \hline
     Solid State Nuclear spin  &
     0.6 \cite{muhonen2015quantifying} & 
     1.12e-4 \cite{laucht2015electrically} - 1.5e-4 \cite{muhonen2015quantifying} &
     1.2e-4 \cite{van2012decoherence}*
     &
     99.6 - \cite{laucht2015electrically} - 99.95 \cite{muhonen2015quantifying}
     &
     89 \cite{jelezko2004observation} - 96 \cite{van2012decoherence}* &
     NO \\
     \hline
     Solid State Electron spin  & 
     1e-3 \cite{nielsen2000quantum} & 
     3.0e-6 \cite{muhonen2015quantifying} - 2.3e-5 \cite{laucht2015electrically} & 
     1.2e-4 \cite{van2012decoherence}* & 
     99.4 \cite{laucht2015electrically} - 99.93 \cite{muhonen2015quantifying} & 
     89 \cite{jelezko2004observation} - 96 \cite{van2012decoherence}* & NO \\
     \hline
     Quantum Dot &
     1e-6 \cite{petta2005coherent,nielsen2000quantum} - 4e-4 \cite{kawakami2016gate} &
     1e-9 \cite{nielsen2000quantum} - 2e-8 \cite{nichol2017high} &
     1e-7 \cite{veldhorst2015two} & 
     98.6 \cite{nichol2017high} - 99.9 \cite{yoneda2018quantum} &
     90 \cite{nichol2017high} &
     NO \\
     \hline
     %Photon &     5e-8 \cite{sipahigil2016integrated} - 1e-5 \cite{nielsen2000quantum} &      &     &     &     98.85 \cite{qiang2018large} &     YES \\  \hline
     NMR &
     16.7 \cite{jiang2018experimental} &
     2.5e-4 \cite{jiang2018experimental} - 1e-3 \cite{xin2018nmrcloudq} &
     2.7e-3 \cite{jiang2018experimental} - 1.0e-2 \cite{xin2018nmrcloudq} &
     98.74 \cite{xin2018nmrcloudq} - 99.60 \cite{jiang2018experimental} &
     98.23 \cite{xin2018nmrcloudq} - 98.77 \cite{jiang2018experimental} & NO \\
     \hline
    
\end{tabular}
}
\caption{Metrics for various quantum technologies. {* Nuclear/Electron Hybrid}}
\label{tab:metrics}
\end{table}

\begin{figure}[h]
    \centering
   % \begin{subfigure}{.45\textwidth}
        \centering
        \includegraphics[scale=0.45]{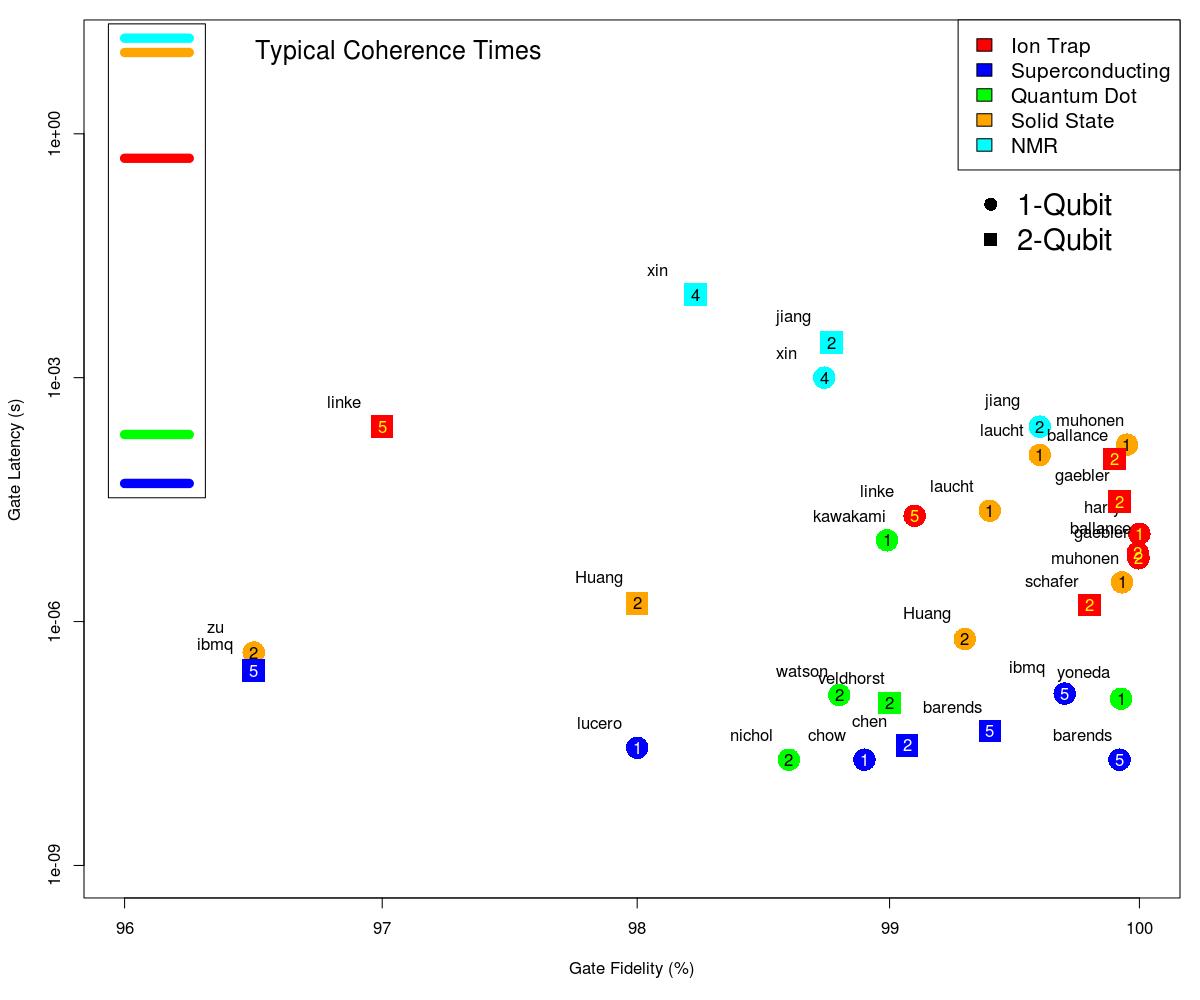}
        \caption{Experimental latencies and fidelities of 1- and 2-qubit gates for different technologies. Approximate coherence times are shown in the inset for comparison. Higher fidelity and lower latency (relative to coherence time) are desirable. Numbers in points indicates total number of qubits in system. Technologies included are Ion Trap \cite{schafer2018fast,ballance2016high,gaebler2016high,harty2014high,linke2017experimental}, Superconducting \cite{rigetti2012superconducting,fu2017experimental,lucero2008high,chow2009randomized,larsen2015semiconductor,ibmexperience,chen2014qubit,barends2014superconducting,chen2014qubit}, Quantum Dot \cite{nichol2017high,yoneda2018quantum,veldhorst2015two,watson2018programmable}, Solid State \cite{muhonen2015quantifying,laucht2015electrically,zu2014experimental,huang2018fidelity}, and NMR \cite{jiang2018experimental,xin2018nmrcloudq}}
        \label{fig:latvfid}
    %\end{subfigure}
    %\begin{subfigure}{.45\textwidth}
    %    \centering
    %    \includegraphics[scale=0.5]{CoherenceBarPlot.jpg}
    %    \caption{Typical coherence times}
    %    \label{fig:CoherenceBarplot}
    %\end{subfigure}
    %\caption{Ion Trap \cite{schafer2018fast,ballance2016high,gaebler2016high,harty2014high} Superconducting \cite{rigetti2012superconducting,fu2017experimental,lucero2008high,chow2009randomized}, Quantum Dot \cite{nichol2017high,yoneda2018quantum,veldhorst2015two}, Donor Spin \cite{muhonen2015quantifying},NMR \cite{jiang2018experimental,xin2018nmrcloudq}}
    %\label{fig:graph}
\end{figure}

Quantitative measures of each metric for various technologies are shown in Table \ref{tab:metrics}. We emphasize that results reported here are for a high level perspective on different technologies and do not directly predict future scalability. For each metric there is typically a wide range, as the performance depends significantly on the specific implementation as well as the underlying technology. As devices are under development, performances can change rapidly. There is a substantial trade-off space as all metrics are inter-related. A notable feature is that coherence times and latencies are, unfortunately so, inversely related. Isolation from the environment increases coherence times, yet makes interaction with qubits for logic gates more difficult, which increases the latency for gate operations \cite{nielsen2000quantum}. Thus, there is a coherence-controllability trade-off \cite{yoneda2018quantum}. A high level evaluation, suggested in \cite{nielsen2000quantum}, is to divide the coherence time by the latency of individual quantum operations. Doing so gives a rough estimate on the number of quantum operations that can be performed, and therefore the size of the algorithms one would be able to compute. This is also referred to as the quality factor \cite{yoneda2018quantum}. This is an important consideration, as clearly a computer which does not have a coherence time significantly longer than its gate latency will not be scalable. However, satisfying this constraint, even if quite significantly, does not inherently mean scalability. Numerous single qubit experiments have demonstrated large coherence-latency ratios \cite{yoneda2018quantum,fu2017experimental,lucero2008high,chow2009randomized,muhonen2015quantifying,larsen2015semiconductor,harty2014high}. However, these systems are untested beyond a single qubit and the effect of scaling on these coherence times and latencies is not clear. The fidelity of individual quantum gates on single qubits is a poor indicator of the fidelity of the same operations applied to larger systems \cite{erhard2019characterizing,dehollain2016optimization}. Coherence-latency ratio as a metric also does not account for gate fidelities. It is unimportant how many gates can be performed within the coherency window if the operations themselves are destroying the quantum state beyond repair. Therefore, gate fidelities must be high enough either to enable effective error correction, or finish the algorithm with a reasonably high chance of correctness. While no useful operations can be performed with a system with such few qubits (most systems do not contain a universal set of gates), such advances are still important steps forward in enabling larger scale quantum computation. Viewing the problem as a whole, a system should have \textit{enough} qubits to hold the  required data and \textit{enough} gate fidelity and coherence-latency ratio to perform all required operations. Figure \ref{fig:latvfid} shows the results of fidelity, absolute latency, and qubit counts of quantum gate experiments for various technologies, along with typical coherence times for each. Figure \ref{fig:latvfid} is useful for insight but it should be noted that it does not represent a fair comparison between all experiments. The experimental setup and goals for each can vary significantly. For example, \cite{xin2018nmrcloudq} is a 4 qubit computer intended for cloud services, whereas \cite{harty2014high} is a single qubit experiment specifically attempting to achieve high gate fidelity. The quantum dot in \cite{kawakami2016gate} appears to have an excessively high latency. However, it has a coherence time significantly longer than the other experiments. Coherence times are highly variable and are not the focus of the experiments shown, so only a rough estimate is shown in the figure. It it noteworthy that other operations, such as measurement, have associated fidelities as well \cite{quantumCS}. However, the requirements are most strict for the gate operations \cite{dehollain2016optimization}.

Despite not being included in Figure \ref{fig:latvfid}, photonic \cite{qiang2018large,santagati2017silicon} and topological quantum computing are prominent approaches. However, they are less similar to other technologies, which makes them difficult to compare on the supplied metrics. Photonic gates are normally applied when a photon passes through some medium. While the fidelity of such gates is natural to consider, the latency is not. It could be considered as the time it takes the photon to pass through a certain region. However, this doesn't provide the same perspective as latency does for other technologies, where the latency is the length of the applied pulse. While there are proposals for topological computers \cite{karzig2017scalable} and physical experiments are creating progress towards demonstration \cite{aasen2016milestones,yang2019spin}, topological qubits have yet to be fully tested. Likely, their operations will take longer than other technologies. Their potential lies in their possible near immunity to traditional sources of quantum noise. This is a very appealing feature, given error is the main obstacle for quantum computing.

Of the listed technologies, ion traps and photon based computers have mobile qubits. The mobility of ion traps is likely to facilitate in entangling and easy communication between qubits, whereas the main benefit of the mobility of photons is likely going to be in the creation of long distance quantum networks. The other technologies have stationary qubits. Mobility in and of itself appears to be an attractive quality as stationary qubits are often restricted to nearest neighbor only communication. For example, mobile ions can be moved to enable entanglement between distant qubits, whereas superconductors must rely on sequences of swap gates to move quantum states. (Alternatively, superconductors can couple over longer distance superconducting transmission lines, however this demands a number of connections which is hard to deliver in 2D layouts \cite{brecht2016multilayer}). This would seem to be an all out advantage for ion traps, however movement also increases risk of decoherence \cite{isailovic2006interconnection,isailovic2004datapath,balensiefer2005evaluation}. Mobility is an advantage only if it provides faster or higher fidelity entanglement between distant qubits. A design could make use of multiple qubit technologies, where stationary qubits can communicate long distance via a network of mobile qubits \cite{nickerson2013topological,sipahigil2016integrated,humphreys2018deterministic}. Taking this to the extreme, all data qubits are stationary and all communication occurs over the network. Such a configuration is a distributed system, where scalability may be possible through modular design of the individual nodes \cite{van2006distributed,nickerson2013topological} and increasing the qubit count just involves adding more nodes. This is a similar concept to that of a quantum internet, but with a focus on shorter distance communication to create a unified system. 

{While these physical properties should be considered when attempting to scale a quantum computer, the authors of \cite{blume2019metrics} warn against establishing standard metrics and benchmarks. It is important to remember that quantum computing is an emerging field. This means progress will be exploratory, rather than incremental improvements on existing techniques. There are many competing approaches, each with unique challenges. Since it is not yet known which approaches will be the most effective, it is not possible to create universally useful metrics or standard benchmarks. Attempting to prematurely do so could actually be detrimental to the field. Research may be directed towards improving results on non-representative benchmarks, which are poor indicators of progress. }

\subsubsection{Engineering}
Beyond the physical properties of the qubits themselves, there are many other considerations which will determine how scalable a technology is. These considerations fall under the broad categorization of engineering. 
\begin{itemize}
    \item \textbf{Topology} is the connectivity between qubits. It will be greatly affected by the mobility of the qubits. Some technologies have immobile qubits and only allow nearest neighbor communication. In such cases the topology is the chosen connections. %Technologies with mobile qubits may be able to interact all qubits with each other, however there are likely to be limitations with large numbers of qubits \cite{linke2017experimental}.
    \item  \textbf{Maturity} represents how developed a technology is and how experienced the community is at building it. While not a characteristic of the technology itself, maturity can have a significant impact on the focus it is given.
    \item \textbf{Fabrication} is an important consideration when extending from experiment to commercial production. Potentially thousands or millions of qubits will have to be reliably integrated. Such designs will require chips with large and complex circuits, which will have to be mass produced.
    \item \textbf{Control and Integration} can become difficult with large numbers of qubits. Integrating control circuitry along with many qubits can be difficult with the extreme noise sensitivity of quantum information. Analog signals are required to drive quantum operations, which can be difficult to deliver quickly and accurately at a large scale.
\end{itemize}

Entangling qubits is an essential component of quantum computing. Doing so allows for the creation of higher dimensional quantum states, which gives quantum computing its advantage over classical computers. The topology of a quantum computer affects the ease at which qubits can communicate and become entangled. As superconducting computers have immobile qubits, the topology determines which qubits can interact. Long distance communication will have to resort to swap gates, where the quantum states of neighboring qubits are swapped. Using this approach increases the number of gates required. The increase in the number of gates can be as much as proportional to the number of qubits \cite{cheung2007translation, linke2017experimental}. However, it may be sublinear or even logarithmic, depending on the design \cite{cowtan2019qubit}. Technologies with mobile qubits can potentially interact all qubits with any other, though the topology will determine if and if so, with what ease.  As systems get larger, local interactions will be insufficient and longer distance communication will have to be considered \cite{quantumCS}. IonQ's ion trap computer can hold 160 qubits, but has only performed full quantum operations on 11 of them \cite{ionqphysicsworld}. The effects of topology were highlighted in \cite{linke2017experimental} where they compared the performance of two different quantum computers on a variety of benchmarks. One computer was a 5-qubit superconducting computer from IBM \cite{ibmexperience}, organized in a star topology, one center qubit connected to four surrounding qubits. The other was their fully-connected 5-qubit ion trap computer. The probability of success (measuring the correct output) dropped significantly on the IBM computer on benchmarks that required substantial amounts of interaction between non-connecting qubits. As larger quantum algorithms are implemented, many more qubits will need to be simultaneously entangled. The topology of future many-qubit computers will likely be critical to their performance.

Maturity is an important metric in these initial days of quantum computing. The full potential of each platform is currently unknown, but the effectiveness of some is more known than others. For example, superconducting computers have dominated the early years and many are in existence today. They are the main focus of multiple successful companies, such as IBM, Google, and start-ups D-Wave and Rigetti. Ion-trap computers also received much attention in the early years and have had notable advances recently. Superconducting and Ion-Trap computers are considered to be at the highest maturity level \cite{linke2017experimental}, given that they have been used to build fully-programmable multi-qubit machines with high level interfaces, something that is yet to be done for other technologies. In contrast, topological quantum computers are still largely unproven. While topological computers may be more successful long term, superconducting and ion trap computers have a lower investment risk. 
 
Fabrication techniques for classical computers are extremely sophisticated and have undergone decades of intense engineering and optimizations. Quantum computers can be quite different in structure and introduce different considerations. If quantum computing comes to prevalence, such chips will have to be mass produced efficiently \cite{brecht2016multilayer}. The degree to which traditional fabrication techniques benefit the construction of quantum computers depends on how similar the underlying technologies are. Overall, semiconductor and superconductor technologies are currently the most compatible \cite{pillarisetty2018qubit}. For example, superconducting computers with qubits based on Josephson junctions can be built with traditional circuit design and produced with well established lithographic techniques \cite{brecht2016multilayer}. However, it is noteworthy that in the near term the development and fabrication process is different for quantum computers \cite{NQIA}. Large classical chip manufacturers must focus on mass production and scaling to smaller node sizes. Quantum computing currently doesn't have such demand. Rigetti, a maker of superconducting quantum chips, is a small company that manages to do quite a bit with a limited budget, in large part because they do not need to mass produce and can use outdated fabrication technology \cite{NQIA}. 

There is a significant gap between high-level quantum programs and the physical realization of quantum gates. While the bulk of useful computation happens within the qubits themselves, classical control is responsible for issuing the quantum instructions, driving the instruments that implement quantum gates, and making decisions for error correction procedures. As more qubits are integrated, the complexity of this problem will increase. Hence, quantum chips will require significant amounts of classical circuitry and computing power. Relative to other sub-fields in quantum computing, there is little research applied to this task \cite{fu2017experimental,fu2018eqasm}.  There are two main considerations for this process to achieve scalability. One is simply the ability to efficiently schedule and implement the gates. While classical computers run considerably faster than the operations performed in quantum computing, this is still a difficult task. Most quantum technologies require analog signals to drive the operations. Quantum operations are time sensitive, requiring precise timing and fast feedback \cite{fu2017experimental}. For superconducting computers, a delay in a pulse on the order of a couple nanoseconds can result in incorrect operation. Ion-traps require accurate and stable delivery of laser beams \cite{linke2017experimental}. Scheduling and delivering these analog signals accurately and with a high degree of precision will be challenging for systems with thousands or millions of qubits. The currently used experimental methods for generating these analog signals are typically not scalable. The solution will likely be architectural, rather than software based, due to the strict latency requirements \cite{fu2017experimental}. The authors of \cite{fu2017experimental} believe the solution is to separate the system stack into upper and lower levels. The upper level can be entirely classical and non-deterministic. The classical upper level can then send commands to a deterministic lower level, which converts classical commands into analog pulses. This allows for a separation of classical, logical control and analog waveform generation. Additionally, the analog components can be abstracted away, allowing classical control to be applied to different underlying technologies. The second major consideration for scalability is performing these control operations, and integrating the necessary machinery, without introducing too much noise. This poses quite a challenge as quantum information is extremely fragile and typically is kept close to absolute zero, well below \SI{1}{\kelvin} \cite{patra2018cryo}. Commonly, the classical control operates at room temperature. When there are many qubits, routing from each qubit, through this temperature differential, to the controller becomes unmanageable. Superconducting computers require large numbers of current carrying wires which can be difficult to route \cite{linke2017experimental,barends2014superconducting,corcoles2015demonstration,riste2015detecting,ofek2016extending,takita2016demonstration}. A possible solution is to move the control circuitry closer to the qubits and operate at very low temperatures, which can enhance compactness and reliability \cite{patra2018cryo,asaad2016independent,chow2014implementing,kawakami2014electrical,kawakami2016gate,veldhorst2015two,veldhorst2014addressable,dicarlo2009demonstration}. Due to limitations in cooling power, these circuits typically have to operate at a warmer temperature than the qubits, around \SI{4}{\kelvin}. Such circuits need to be designed not only to generate very little noise, but deliver the performance required for the previously mentioned complex control operations. CMOS circuits appear to be well suited for this purpose \cite{patra2018cryo,ekanayake2010characterization,hornibrook2015cryogenic,homulle2016cryocmos,charbon2016cryo,sebastiano2017cryo}.  FPGA based control schemes have also been proposed and tested, including being operated at very low temperatures allowing them to be placed relatively near the qubits \cite{hornibrook2015cryogenic,homulle2017reconfigurable}.  Another challenge to integrating and controlling large numbers of qubits comes from the qubits themselves. When close together, there can be unwanted couplings (referred to as "cross-talk" between qubits which can lead to the mixture of quantum states or decoherence \cite{brecht2016multilayer}. A stable design should be able to both keep qubits well isolated and strongly couple them, when desired. A number of designs for various technolgoies have been proposed attempting to solve these concerns. A 3-dimensional layout using multiple layers using a process called "micromachining" was proposed for superconducting chips in \cite{brecht2016multilayer}. A 3D integration scheme was demonstrated in \cite{das2018cryogenic}, where there were able to connect multiple superconducting chips and also chips for read out and interconnection. A multi-layer ion trap was constructed with 3-dimensional microwave circuitry in \cite{bautista2018multilayer}, where they claim an arbitrary number of layers is possible. A design for the integration of ion trap and superconducting qubits into a hybrid system is proposed in \cite{de2016experimental}.

\subsection{Going Forward}
While much progress has been made, there is still a long ways to go. Quantum remains an emerging field and there are many open problems to pursue. 

 Devices with significantly lower error rates are essential to building large-scale computers. The overhead of quantum error correcting codes is substantial and highly dependent on underlying error rates \cite{oskin2002practical}. The error rate needs to be sufficiently low so that the addition of more qubits and longer coherence requirements does not cause an intolerable need for error correction. All physical quantum technologies are improving and only time will tell if reasonable thresholds are met. 

Architectures that are scalable with the number of qubits are required to build larger computers. Noise becomes more of an issue with increasing size and the architecture has to be built to handle this. Designs must be capable of implementing complex control operations and performing efficient error correction, all while not introducing excessive noise themselves. Additionally, the chips they require must be mass producible. 

Algorithms that make use of noisy quantum computers are in high demand. Clever algorithms can lower the physical and engineering requirements in order to build useful quantum  machines. Such algorithms are useful not only because they provide applications and benchmarks for current quantum computers, but they provide insight into the potential of quantum computing and motivate efforts towards ever larger scale implementations.

\pagebreak
\begin{table*}[htbp]
    \small
    \centering
    \caption{\textbf{At a Glance:} Summary of different quantum technologies.}
    \label{tab:my_label}
    \resizebox*{1.0\linewidth}{!}{
    \begin{tabular}{|p{2cm}|p{4cm}|p{4cm}|p{4cm}|}
        \hline
         Technology & Basic Description & Advantages & Disadvantages/ Challenges  \\
         \hline
         Superconductors &
         Charge, Current, or Energy of superconducting circuit \cite{flamini2018photonic}
         &
          High compatibility with existing fabrication techniques \cite{brecht2016multilayer};
          Electronic control \cite{brecht2016multilayer}; Easy coupling \cite{gonzalez2018solid}; Mature technology \cite{quantumCS}
         
         &
         Large footprint \cite{quantumCS}
         
         \\
         \hline
         Quantum Dot & 
         Semiconductor particles a few nanometers in size. Can be constructed in semiconductors with controllable numbers of electrons, including 0 to 1 \cite{eriksson2004spin}. The spin of these electrons can be used as qubits. &
         Potential scalability with well established fabrication techniques \cite{yoneda2018quantum,quantumCS,kawakami2016gate}
         ; All electrical operation, including electrically controllable spin-spin coupling \cite{ess1,veldhorst2015two,watson2018programmable,maurand2016cmos,vandersypen2017interfacing}
         ; Potential high density \cite{watson2018programmable,maurand2016cmos,vandersypen2017interfacing}
         & 
        Decoherence due to electrostatic fluctuation \cite{russ2018high,twelve,petta2005coherent}
         \\
        \hline
        Ion Trap &
        Individual atom held in a vacuum via an electromagnetic trap generated by surrounding electrodes. Laser pulses perform gate operations \cite{nielsen2000quantum}.
        &
        Long coherence times \cite{de2016experimental}; Mature technology \cite{quantumCS}
        
        &
        Fluctuating electric and magnetic fields push on the ions, causing decoherence \cite{nielsen2000quantum}
 
        \\
        \hline
        Photons &
        Information encoded in polarization, orbital angular momentum, number (0/1 photons), or time \cite{flamini2018photonic}. Interact with phase shifters, beamsplitters, optical media, and photodetectors \cite{kok2007linear,nielsen2000quantum}. 
        &
        Lack of interaction with environment reduces decoherence \cite{kok2007linear,nielsen2000quantum,flamini2018photonic}
        ; Mobility makes them ideal for quantum network communication \cite{flamini2018photonic}
        ; Built on silicon infrastructure \cite{buildphoton,qiang2018large}
        
        &
        Difficult to interact with to perform gates \cite{kok2007linear,nielsen2000quantum}
        ; Difficult to interact them with each other
        ; Kerr media is absorptive and scatters light \cite{nielsen2000quantum}
        ; Requires precise control of large circuits of linear optical components \cite{qiang2018large}; Lack of single photon sources \cite{quantumCS}

        \\
        \hline
        Solid-State Spin 
        &
        Nuclear or electron spin of donor atoms in a semiconductor or NV centers in diamond. 
        &
        Highly coherent \cite{sigillito2017all}; potential CMOS compatibility \cite{morse2017photonic}
        &
        High precision fabrication requirements \cite{gonzalez2018solid} ; Slower operations for nuclear spins \cite{sigillito2017all} 

        \\
       % \hline
        %NMR &
        %Weakly polarized macroscopic ensembles of spins manipulated by nuclear magnetic resonance spectroscopy \cite{cory1998nuclear}
        %&
        %Technology is mature \cite{xin2018nmrcloudq}
        %&
        
        %\\
        \hline
        Topological
        &
        Non-abelian anyons can be created in superconductors and topological insulators \cite{pachos2012introduction}. Gates performed by braiding the anyons \cite{nayak2008non} or by performing measurements \cite{bonderson2008measurement}. %To braid anyons, 1-D semiconductor wires can be driven into a topological phase \cite{lutchyn2010majorana,oreg2010helical} and then voltages are applied to local electrodes to create, move, and destroy the quasi-particles \cite{alicea2011non}.
        &
        Hardware level resistance to error \cite{pachos2012introduction}
        &
        Less mature than other approaches; Hard to engineer \cite{quantumCS}
        
        \\
        \hline
    \end{tabular}
    }
    
\end{table*}
\clearpage